\documentclass[10pt,journal,compsoc]{IEEEtran}
\ifCLASSOPTIONcompsoc
  \usepackage[nocompress]{cite}
\else
  \usepackage{cite}
\fi

\usepackage{graphicx}
\usepackage{subfigure}
\usepackage{multirow}
\usepackage{multicol}
\usepackage{amsfonts}
\usepackage[ruled]{algorithm2e}
\usepackage{algpseudocode}
\usepackage{bbding}
\usepackage{flushend}
\ifCLASSINFOpdf
\else
\fi

\begin{document}

\title{On Benchmarking the Capability of Symbolic Execution Tools with Logic Bombs}

\author{Hui~Xu,
        Zirui~Zhao,
        Yangfan~Zhou,
        and~Michael~R.~Lyu,~\IEEEmembership{Fellow,~IEEE}
\IEEEcompsocitemizethanks{\IEEEcompsocthanksitem H. Xu, and M. R. Lyu are with Shenzhen Research Institute and
Dept. of Computer Science \& Engineering, The Chinese University of Hong Kong. Email: {hxu, lyu}@cse.cuhk.edu.hk
\IEEEcompsocthanksitem Z. Zhao is with The Chinese University of Hong Kong and The University of Science and Technology of China
\IEEEcompsocthanksitem Y. Zhou\Envelope \enskip is with School of Computer Science, Fudan University. Email: zyf@fdu.edu.cn
}
\thanks{Manuscript submitted to IEEE Transactions on Dependable and Secure Computing.}}

\IEEEtitleabstractindextext{%
\begin{abstract}
Symbolic execution now becomes an indispensable technique for software testing and program analysis.  There are several symbolic execution tools available off-the-shelf, and we need a practical benchmark approach to learn their capabilities.  Therefore, this paper introduces a novel approach to benchmark symbolic execution tools in a fine-grained and efficient manner.  In particular, our approach evaluates the performance of such tools against the known challenges faced by general symbolic execution techniques, such as floating-point numbers and symbolic memories.  To this end, we first survey related papers and systematize the challenges of symbolic execution.  We extract 12 distinct challenges from the literature and categorize them into two categories: \textit{symbolic-reasoning challenges} and \textit{path-explosion challenges}.  Then, we develop a dataset of logic bombs and a framework to benchmark symbolic execution tools automatically.  For each challenge, our dataset contains several logic bombs, each of which is guarded by a specific challenging problem.  If a symbolic execution tool can find test cases to trigger logic bombs, it indicates that the tool can handle the corresponding problems.  We have conducted real-world experiments with three popular symbolic execution tools: KLEE, Angr, and Triton.  Experimental results show that our approach can reveal their capabilities and limitations in handling particular issues accurately and efficiently.  The benchmark process generally takes only dozens of minutes to evaluate a tool.  We release our dataset on GitHub as open source, with an aim to better facilitate the community to conduct future work on benchmarking symbolic execution tools.

\end{abstract}

}

\maketitle


\ifCLASSOPTIONcompsoc
\IEEEraisesectionheading{\section{Introduction}\label{sec:introduction}}
\else
\section{Introduction} \label{sec:introduction}
\fi

Symbolic execution is a popular technique for software testing and program analysis~\cite{king1976symbolic}.  It has achieved rapid development in the last decade with several open-source symbolic execution tools available, such as KLEE~\cite{cadar2008klee} and Angr~\cite{shoshitaishvili2016state}.  Current methods for evaluating symbolic execution tools generally rely on the achieved code coverage or the number of bugs detected in real-world programs (\textit{e.g.},~\cite{cadar2008klee,ramos2015under}).  Nevertheless, such metrics may fluctuate with different types of programs being analyzed, and they cannot manifest the detailed capabilities of a symbolic execution tool.  This paper, therefore, aims to propose a fine-grained benchmark approach for symbolic execution tools which is less sensitive to targeting programs.

In particular, we observe that there are some common challenging problems which symbolic execution tools may not handle well, such as floating-point numbers~\cite{quan2016hotspot} and loops~\cite{cadar2013symbolic}.  They are the determinant factors of the code coverage that a symbolic execution tool can achieve.  Therefore, we propose to develop a benchmark approach based on these known challenges.  To elaborate, we can employ each challenge as an evaluation metric.  In this way, our approach can give more meaningful information concerning the capability of a symbolic execution tool.  Moreover, the benchmark result is unbiased as it does not depend on particular programs for analysis.

To this end, we first conduct a systematic survey on the challenges of symbolic execution.  This step is essential for our benchmark approach to embrace as many distinct challenges as possible.  We categorize existing challenges into two categories: \textit{symbolic-reasoning challenges} and \textit{path-explosion challenges}.  Symbolic-reasoning challenges attack the core symbolic reasoning process, where they may pose problems to a symbolic execution tool in generating incorrect test cases for particular control flows.  These challenges include symbolic variable declarations, covert propagations, parallel executions, symbolic memories, contextual symbolic values, symbolic jumps, floating-point numbers, buffer overflows, and arithmetic overflows.  Path-explosion challenges introduce too many possible control flows to analyze, which may cause a symbolic execution tool starving the computational resources or spending very long time on exploring the paths.  Not only large-sized programs but also small-sized programs can cause path-explosion issues, as long as they include complex routines, such as external function calls, loops, and crypto functions.  With our approach, consequently, all existing challenges discussed in the literature can be well categorized.

\begin{figure*}[t]
\centering
\includegraphics[width=0.96\textwidth]{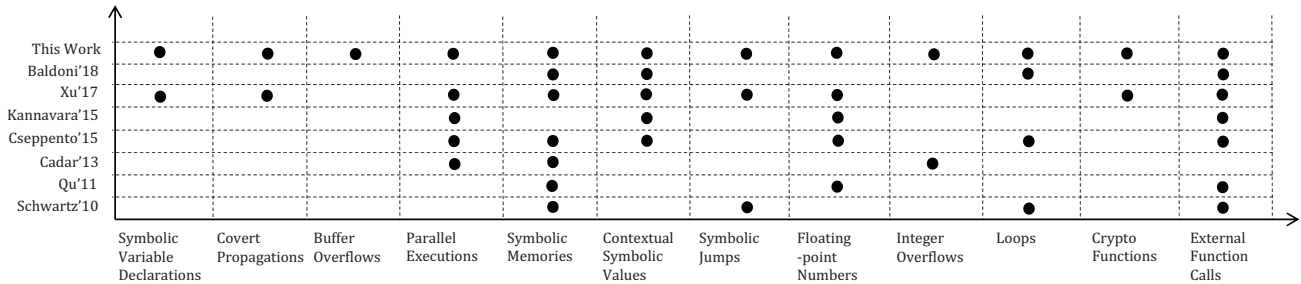}
\caption{The challenges of symbolic execution discussed in the literature.  The detailed paper references are Schwartz'10~\protect\cite{schwartz2010all}, Qu'11~\protect\cite{qu2011case}, Cadar'13~\protect\cite{cadar2013symbolic}, Cseppento'15~\cite{cseppento2015evaluating}, Kannavara'15~\protect\cite{kannavara2015challenges}, Quan'16~\protect\cite{quan2016hotspot}, Xu'17~\protect\cite{xu2017concolic}, and Baldoni'18~\protect\cite{baldoni2018survey}.}
\label{fig:literature}
\end{figure*}

Next, we develop an accurate and efficient approach to benchmark the capability of symbolic execution tools with respect to each of the challenges.  To this end, we cannot employ real-world programs for testing because they are too complicated and any challenges could cause a failure.  Moreover, symbolic execution itself is not very efficient, and benchmark with real-world programs generally takes a long time.  We tackle the problem by designing small programs embedded with logic bombs.  A logic bomb is an artificial code block that can only be executed when certain conditions are met.  We create such logic bombs that can only be triggered when a challenging problem is solved.  The benefits are two folds.  By keeping each logic bomb as small as possible, our evaluation result would be less likely affected by other unexpected issues.  Also, employing small programs can shorten the required symbolic execution time and provides efficiency.

Following this method, we have designed a dataset of logic bombs covering all the challenges and a framework to run benchmark experiments automatically.  For each challenge, our dataset contains several logic bombs with different problem settings or different levels of hardness, such as a one-leveled arrays or a two-leveled arrays for the symbolic memory challenge.  Our framework employs the dataset of logic bombs as evaluation metrics.  It firstly parses the logic bombs and compiles them to object codes or binaries; then, it directs a symbolic execution tool to perform symbolic execution on the logic bombs in a batch mode; and finally, it verifies the generated test cases and produces reports.  We release our dataset and framework tools on GitHub.

We have conducted real-world experiments to benchmark three prevalent symbolic execution tools, KLEE, Triton, and Angr.  Although these tools adopt different implementation techniques, our framework can adapt to them with only a little customization.  The benchmark process for each tool generally took dozens of minutes.  Experimental results show that our benchmark approach can reveal their capabilities and limitations efficiently and accurately.  In summary, Angr has achieved the best performance with 21 cases solved, which is roughly one third of the total logic bombs; KLEE solved nine cases; and Triton only solved three cases.  We manually checked the reported solutions and confirmed that they are all correct and nontrivial.  Besides, the results also demonstrate some interesting findings about these tools.  For example, Angr only supports one-leveled arrays but not two-leveled arrays; Triton does not even support the \texttt{atoi} function.  Most of our findings are unavailable from the tool websites or existing papers, which further justifies the value of our benchmark approach.  

The rest of the paper is organized as follows.  We first discuss the related work in Section~\ref{sec:literature} and introduce the preliminary knowledge of symbolic execution in Section~\ref{sec:background}.  Then, we demonstrate our systematic study about the challenges of symbolic execution in Section~\ref{sec:challenges}, and we introduce our benchmark methodology in Section~\ref{sec:method}.  We present our experiments and results in Section~\ref{sec:experiment}.  Finally, we conclude this paper in Section~\ref{sec:conclusion}.

\section{Related Work} \label{sec:literature}

This section compares our work with present papers that either systematize the challenges of symbolic execution or employ the challenges to evaluate symbolic execution tools.  Note that although symbolic execution has received extensive attention for decades, only a few papers include a systematic discussion about the challenges of the technique.  Existing work in this area mainly focuses on employing the technique to carry out specific software analysis tasks (\textit{e.g.,}~\cite{avgerinos2011aeg,ming2015loop,yadegari2015symbolic}), or proposing new approaches to improve the technology concerning particular challenges (\textit{e.g.,}~\cite{xie2009fitness,ciortea2010cloud9,avgerinos2014enhancing}).

The papers that focus on systematizing the challenges of symbolic execution tools include~\cite{qu2011case,cseppento2015evaluating,kannavara2015challenges,banescu2016code}.  Kannavara \textit{et al.}~\cite{kannavara2015challenges} enumerated several challenges that may hinder the adoption of symbolic execution in industrial fields.  Qu and Robinson~\cite{qu2011case} conducted a case study on the limitations of symbolic testing tools and examined their prevalence in real-world programs.  However, none of the two papers provides a method to evaluate symbolic execution tools.  Cseppento and Micskei~\cite{cseppento2015evaluating} proposed several metrics to evaluate source-code-based symbolic execution tools.  But their metrics are based on specific program syntax of object-oriented codes rather than the language-independent challenges.  These metrics are not very general for symbolic execution.  Banescu \textit{et al.}~\cite{banescu2016code} also design several small programs for evaluation.  But their purpose is to evaluate the resilience of code obfuscation transformations against symbolic execution-based attacks.  They do not investigate the capability of symbolic execution but trust KLEE as a state-of-the-art symbolic executor.  Besides, there are several survey papers (\textit{e.g.,}~\cite{schwartz2010all,cadar2013symbolic,baldoni2018survey}) which also include some discussion about the challenges.  But our work provides a more complete list of challenges as shown in Figure~\ref{fig:literature}.

In our previous conference paper~\cite{xu2017concolic}, we have conducted an empirical study with some of the challenges.  This paper notably extends our previous paper with a formal benchmark methodology.  It serves as a pilot study on systematically benchmarking symbolic execution tools in handling particular challenges.  We consequently design a novel benchmark framework based on logic bombs, which can facilitate the automation of the benchmark process.  We further provide a benchmark toolset that can be easily deployed by ordinary users.

\section{Preliminary} \label{sec:background}
This section reviews the underlying techniques of symbolic execution, which is a preliminary for discussing the challenges and the benchmark approach we proposed in this work.

\subsection{Theoretical Basis}
The core principle of symbolic execution is symbolic reasoning.  Informally, given a sequence of instructions along a control path, a symbolic reasoning engine can extract a constraint model and generates a test case for the path by solving the model.

Formally, we can use Hoare Logic~\cite{hoare1969axiomatic} to model the symbolic reasoning problem.  Hoare Logic is composed of basic triples $\{S_1\}I\{S_2\}$, where $\{S_1\}$ and $\{S_2\}$ are the assertions of variable states and $I$ is an instruction. The Hoare triple says if a precondition $\{S_1\}$ is met, when executing $I$, it will terminate with the postcondition $\{S_2\}$.  Using Hoare Logic, we can model the semantics of instructions along a control path as:

$$\{S_0\}I_0\{S_1,\Delta_1\}I_1...\{S_{n-1},\Delta_{n-1}\}I_{n-1}\{S_n\}$$

$\{S_0\}$ is the initial symbolic state of the program; $\{S_1\}$ is the symbolic state before the first conditional branch with symbolic variables; $\Delta_i$ is the corresponding constraint for executing the following instructions, and $\{S_i\}$ satisfies $\Delta_i$.  A symbolic execution engine can compute an initial state $\{S'_0\}$ (\textit{i.e.,} the concrete values for symbolic variables) which can trigger the same control path.  This can be achieved by computing the weakest precondition (\textit{aka} $wp$) backward using Hoare Logic:

$$\{S_{n-2}\} = wp(I_{n-2}\{S_{n-1}\}), \quad s.t. \ \{S_{n-1}\} \ sat \ \Delta_{n-1}$$
$$\{S_{n-3}\} = wp(I_{n-3}\{S_{n-2}\}), \quad s.t. \ \{S_{n-2}\} \ sat \ \Delta_{n-2}$$
$$...$$
$$\{S_1\} = wp(I_{1}\{S_2\}), \quad s.t. \ \{S_{2}\} \ sat \ \Delta_{2}$$
$$\{S_0\} = wp(I_0\{S_1\}), \quad s.t. \ \{S_{1}\} \ sat \ \Delta_{1}$$

Combining the constraints in each line, we can get a constraint model in conjunction normal form: $\Delta_1 \wedge \Delta_2 \wedge ... \wedge \Delta_{n-1}$.  The solution to the constraint model is a test case $\{S'_0\}$ that can trigger the same control path.

Finally, when sampling $\{I_i\}$, not all instructions are useful.  We only keep the instructions whose parameter values depend on the symbolic variables.  We can demonstrate the correctness by expending any irrelevant instruction $I_i$ to $X:=E$, which manipulates the value of a variable $X$ with an expression $E$.  Suppose $E$ does not depend on any symbolic value, then $X$ would be a constant, and should not be included in the weakest preconditions.  In practice, it can be realized using taint analysis techniques~\cite{song2008bitblaze}.

\subsection{Symbolic Execution Framework}
\begin{figure}[t]
\centering
\includegraphics[width=0.46\textwidth]{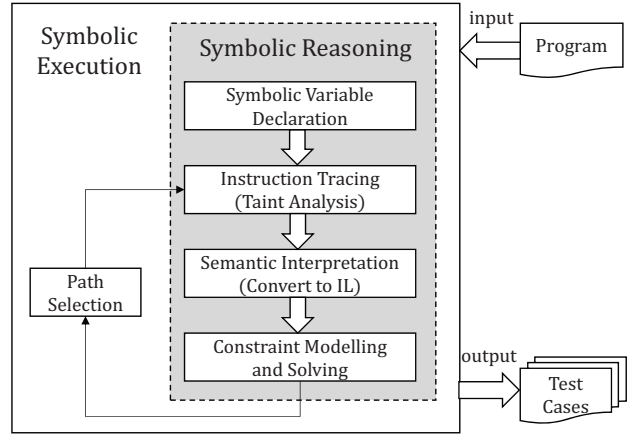}
\caption{A conceptual framework for symbolic execution.}
\label{fig:symexe}
\end{figure}

We demonstrate the conceptual framework of a symbolic execution tool in Figure~\ref{fig:symexe}.  It inputs a program and outputs test cases for the program.  The framework includes a core symbolic reasoning engine and a path selection engine.  

The symbolic reasoning engine analyzes the instructions along a path and generates test cases that can trigger the path.  Based on the theoretical basis of symbolic reasoning, we can divide symbolic reasoning into four stages: symbolic variable declaration, instruction tracing, semantic interpretation, and constraint modeling and solving.  The details are discussed as follows:

\begin{itemize}
\item \textit{Symbolic variable declaration} ($S_{var}$): In this stage, we have to declare symbolic variables which will be employed in the following symbolic analysis process.  If some symbolic variables are missing from declaration, insufficient constraints can be generated for triggering a control path.
\item \textit{Instruction tracing} ($S_{inst}$): This stage collects the instructions along control paths.  If some instructions are missing, or the syntax are not supported, errors would occur.
\item \textit{Semantic interpretation} ($S_{sem}$): This stage translates the semantics of collected instructions with an intermediate language (IL).  If some instructions are not correctly interpreted, or the data propagation are incorrectly modeled, errors would occur.
\item \textit{Constraint modeling and solving} ($S_{model}$): This stage generates constraint models from IL, and then solve the constraint models.  If a required satisfiability modulo theory is unsupported, errors would occur.
\end{itemize}

The path selection engine determines which path should be analyzed in the next round of symbolic reasoning process.  Several favorite strategies include depth-first search, width-first search, random search, \textit{etc}~\cite{baldoni2018survey}.

\begin{table*}[t]
\centering
\caption{A list of the challenges faced by symbolic execution, and the symbolic reasoning stages they attack.}
\label{tab:challenge}
\newcommand{\tabincell}[2]{\begin{tabular}{@{}#1@{}}#2\end{tabular}}
\renewcommand{\multirowsetup}{\centering}
\tabcolsep=0.2cm
\begin{tabular}{|c|c|c|c|c|c|}
\hline
\multicolumn{2}{|c|}{\multirow{2}{*}{\textbf{Challenge}}} & \multirow{2}{*}{\textbf{Idea}} & \multicolumn{3}{|c|}{\textbf{Stage of Error}} \\
\cline{4-6}
 \multicolumn{2}{|c|}{} & & $S_{var}$ & $S_{inst}$ \& $S_{sem}$ & $S_{model}$\\
\hline
\multirow{9}{*}{\tabincell{c}{Symbolic\\-reasoning \\Challenges}} & Sym. Var. Declaration & Contextual variables besides program arguments & \checkmark & \checkmark & \checkmark \\
\cline{2-6}
& Covert Propagations & Propagating symbolic values in covert ways & - & \checkmark & \checkmark \\
\cline{2-6}
& Buffer Overflows & Writing symbolic values without proper boundary check & - & \checkmark & \checkmark \\
\cline{2-6}
& Parallel Executions & Processing symbolic values with parallel codes & - & \checkmark & \checkmark \\
\cline{2-6}
& Symbolic Memories & Symbolic values as the offset of memory & - & \checkmark & \checkmark \\
\cline{2-6}
& Contextual Symbolic Values & Retrieving contextual values with symbolic values & - & \checkmark & \checkmark \\
\cline{2-6}
& Symbolic Jumps & Sym. values as the addresses of unconditional jump & - & - & \checkmark \\
\cline{2-6}
& Floating-point Numbers & Symbolic values in float/double type & - & - & \checkmark \\
\cline{2-6}
& Arithmetic Overflows & Integers outside the scope of an integer type & - & - & \checkmark \\
\hline
\multirow{3}{*}{\tabincell{c}{Path-explosion \\Challenges}} & Loops & Change symbolic values within loops & - & - & -  \\
\cline{2-6}
& Crypto Functions & Processing symbolic values with crypto functions & - & - & - \\
\cline{2-6}
& External Function Calls & Processing sym. values with some external functions & - & - & - \\
\hline
\end{tabular}
\end{table*}

\subsection{Implementation Variations} 
According to the different ways of instruction tracing, we can classify symbolic execution tools into static symbolic execution and dynamic symbolic execution.  Static symbolic execution loads a whole program first before extracting instructions along with a path on the program control-flow graph (CFG).  Dynamic symbolic execution is also known as concolic (concrete and symbolic) execution.  It collects instructions which are actually executed.  In each round, the concolic execution engine executes the program with concrete values to generate instructions.

We may also classify symbolic execution tools into source-code-based symbolic execution and binary-code-based symbolic execution.  In general, we do not perform symbolic reasoning on source codes or binaries directly.  A prior step is to interpret the semantics of the program with an intermediate language (IL).  For source codes, we can translate the code directly with a compiler frontend.  For binaries, we have to lift the assembly codes into IL, which is more difficult.  Their main difference lies in the translation process.

\section{Challenges of Symbolic Execution} \label{sec:challenges}
Based on whether a challenge attacks the symbolic reasoning process, we categorize the challenges of symbolic execution into \textit{symbolic-reasoning challenges} and \textit{path-explosion challenges}.  A symbolic-reasoning challenge attacks the symbolic reasoning process and leads to incorrect test cases generated.  A path-explosion challenge happens when there are too many paths to analyze.  It does not attack a single symbolic reasoning process, but it may starve the computational resources or requires a very long time for symbolic execution.  

Table~\ref{tab:challenge} demonstrates the challenges that we have investigated in this work.  We collect such challenges via a careful survey of existing papers.  The survey scope covers several survey papers about symbolic execution techniques (\textit{e.g.,}~\cite{schwartz2010all,cadar2013symbolic,baldoni2018survey}), several investigations that focus on systemizing the challenges of symbolic execution (\textit{e.g.,}~\cite{cseppento2015evaluating,kannavara2015challenges}), and other important papers related to symbolic execution (\textit{e.g.,}~\cite{godefroid2005dart,xie2009fitness,brumley2011bap,razavi2012concurrent,
cha2012unleashing,davidson2013fie,yadegari2015symbolic,guo2016conc}).

\begin{figure*}[t]
\centering
\subfigure[Symbolic variable declarations.]{
\label{fig:symvar}
\includegraphics[width=0.25\textwidth]{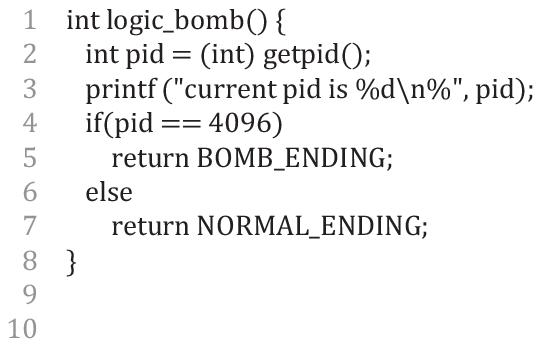}}
\subfigure[Covert symbolic propagations.]{
\label{fig:covpro}
\includegraphics[width=0.46\textwidth]{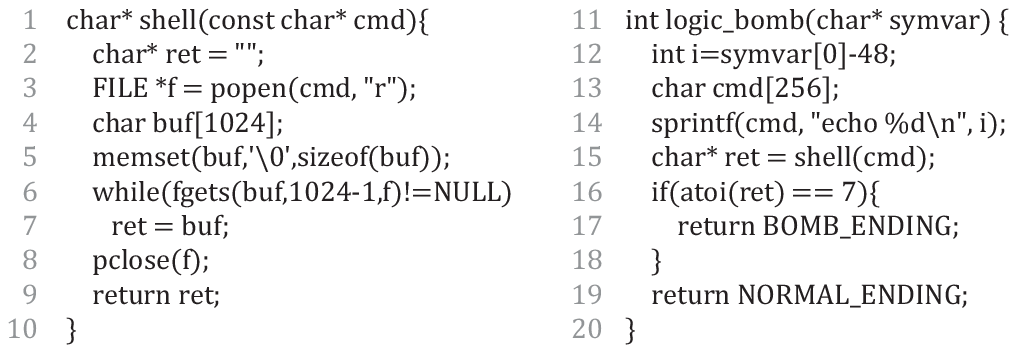}}
\subfigure[Buffer overflows.]{
\label{fig:bufferoverflow}
\includegraphics[width=0.21\textwidth]{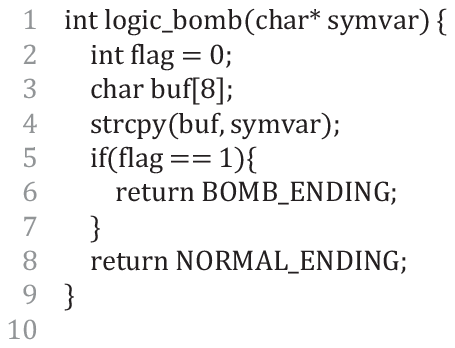}}
\subfigure[Parallel executions.]{
\label{fig:paraprog}
\includegraphics[width=0.56\textwidth]{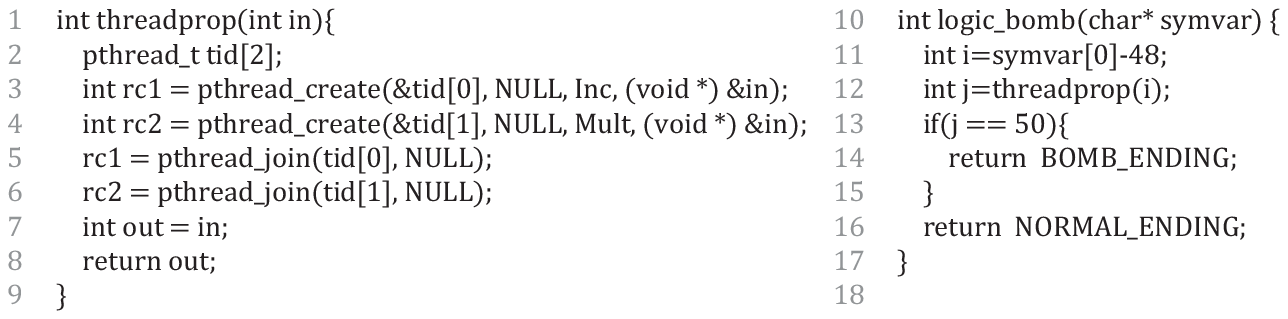}}
\subfigure[Symbolic memories.]{
\label{fig:symarray}
\includegraphics[width=0.20\textwidth]{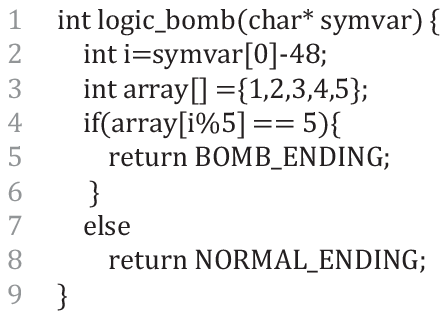}}
\subfigure[Contextual symbolic values.]{
\label{fig:symval}
\includegraphics[width=0.20\textwidth]{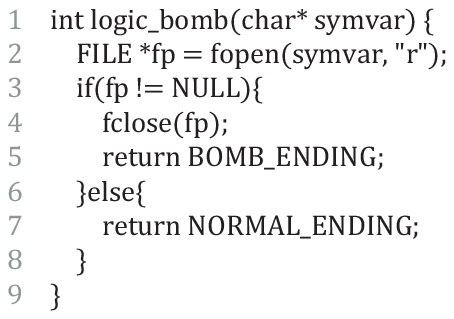}}
\subfigure[Symbolic jumps.]{
\label{fig:symjmp}
\includegraphics[width=0.275\textwidth]{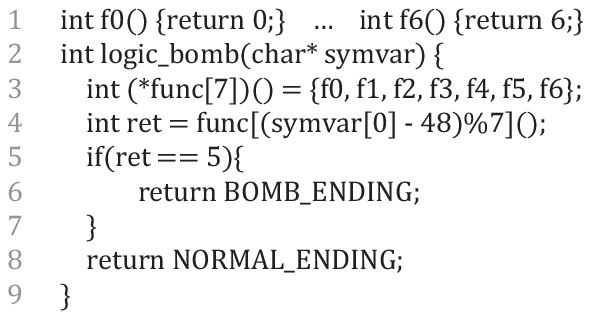}}
\subfigure[Floating-point numbers.]{
\label{fig:floatpoint}
\includegraphics[width=0.22\textwidth]{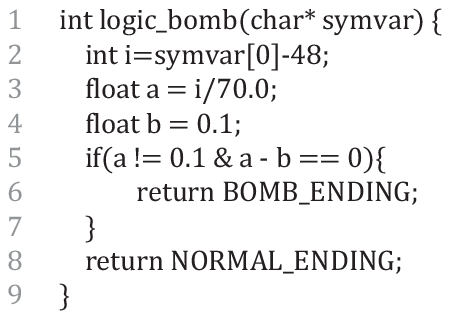}}
\subfigure[Arithmetic overflows.]{
\label{fig:intoverflow}
\includegraphics[width=0.23\textwidth]{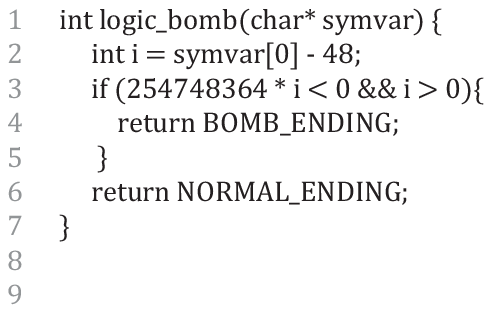}}
\subfigure[External function calls.]{
\label{fig:extfun}
\includegraphics[width=0.21\textwidth]{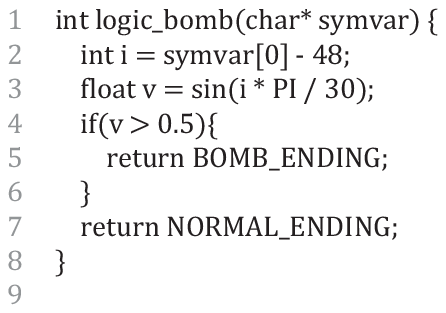}}
\subfigure[Loops.]{
\label{fig:loop}
\includegraphics[width=0.42\textwidth]{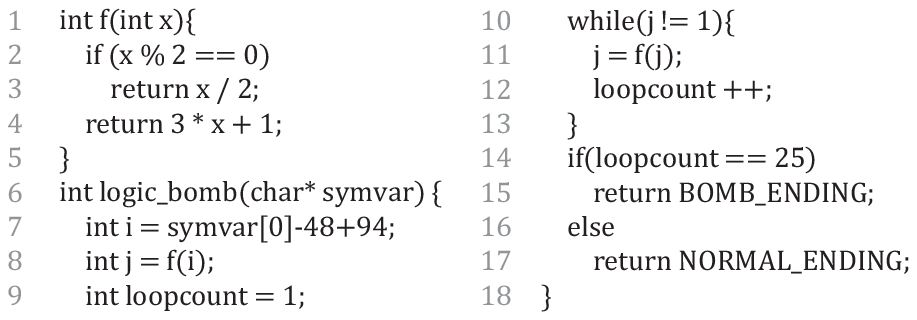}}
\subfigure[Crypto functions.]{
\label{fig:cryptofun}
\includegraphics[width=0.54\textwidth]{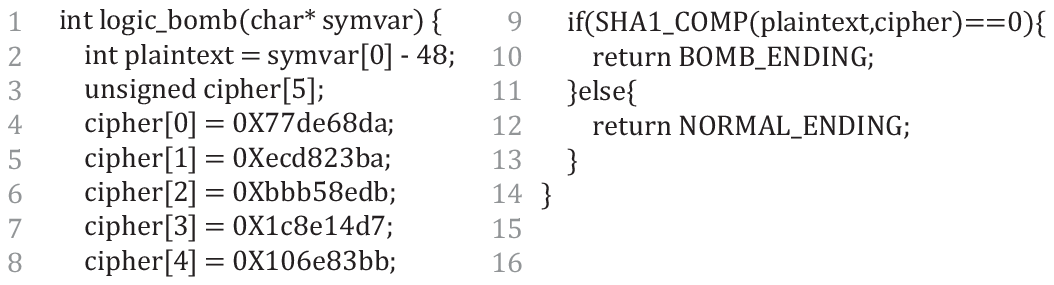}}
\caption{Logic bomb samples with challenging symbolic execution issues.  In each sample, we employ \texttt{symvar} to denote a symbolic variable, and \texttt{BOMB\_ENDING} to denote a macro value indicating a particular program behavior.}
\label{fig:examples}
\end{figure*}

\subsection{Symbolic-reasoning Challenges}
Next, we discuss nine challenges that may incur errors to the symbolic reasoning process.

\subsubsection{Symbolic Variable Declarations}
Test cases are the solutions of symbolic variables subject to constrain models.  Therefore, symbolic variables should be declared before a symbolic analysis process.  For example, in source-code-based symbolic execution tools (\textit{e.g.,} KLEE), users can manually declare symbolic variables in the source codes.  Binary-code-based concolic execution tools (\textit{e.g.,} Triton) generally assume a fixed length of program arguments from stdin as the symbolic variable.  If some symbolic variables are missing from the declaration, the generated test cases would be insufficient for triggering particular control paths.  Since the root cause occurs before symbolic execution, the challenge attacks $S_{var}$.

Figure~\ref{fig:symvar} is a sample with a symbolic variable declaration problem.  It returns a \texttt{BOMB\_ENDING} only when being executed with a particular process id.  To explore the path, a symbolic execution tool should treat \texttt{pid} as a symbolic variable and then solve the constraint with respect to \texttt{pid}.  Otherwise, it cannot find test cases that can trigger the path.

To declare symbolic variables precisely, a user should know target programs well.  However, the task is impossible when analyzing programs on a large scale, \textit{e.g.,} when performing malware analysis.  In an ideal case, a symbolic execution tool may automatically detect such variables which can control program behaviors and reports the solutions accordingly.  To our best knowledge, non-existent tools have implemented the ideal feature.  Instead, they are generally discussed together with other problems related to the computing environment, such as libraries, kernels, and drivers~\cite{chipounov2011s2e}.  In reality, there are several challenges of this work refer to the computing environment, such as contextual symbolic variables, covert propagations, parallel executions, external function calls.  We demonstrate that these challenges are different.

\subsubsection{Covert Propagations}
Some data propagation ways are covert because they cannot be traced easily by data-flow analysis tools.  For example, if the symbolic values are propagated via other media (\textit{e.g.,} files) outside of the process memory, the propagation would be untraceable.  Such propagation methods are undecidable and can be beyond the capability of pure program analysis.  Symbolic execution tools have to handle such cases with ad hoc methods.  There are also some propagations challenging only to certain implementations.  For example, propagating symbolic values via embedded assembly codes should be a problem for source-code-based symbolic execution tools only.  If a symbolic execution tool fails to detect some propagation, the instructions related to the propagated values would be missed from the following analysis.  Therefore, the challenge attacks the stages of $S_{inst}$ and $S_{sem}$.

Figure~\ref{fig:covpro} shows a covert propagation sample.  We define an integer \texttt{i} and initiate it with the value of a symbolic variable (\texttt{symvar}).  So \texttt{i} is also a symbolic variable.  We then propagate the value of \texttt{i} to another variable (\texttt{ret}) through a shell command (\texttt{echo}), and let \texttt{ret} control the return value.  To find a test case which can return the \texttt{BOMB\_ENDING}, a symbolic execution tool should properly track or model the propagation incurred by the shell command.

\subsubsection{Buffer Overflows}
Buffer overflow is a typical software bug that can bring security issues.  Due to insufficient boundary check, the input data may overwrite adjacent memories.  Adversaries can employ such bugs to inject data and intentionally tamper the semantics of the original codes.  Buffer overflows can happen in either stack or heap regions.  If a symbolic execution tool cannot detect the overflow issues, it would fail to track the propagation of symbolic values.  Therefore, buffer overflow involves a particular covert propagation issue.  Source-code-based symbolic execution tools are prone to be affected by buffer overflows because the stack layout of a program only exists in assembly codes, which may vary for particular platforms.  Therefore, such tools cannot model the stack information with source codes only.  In contrast, binary-code-based symbolic execution tools should be more potent in handling buffer overflow issues because there can simulate actual memory layouts.  However, even if these tools can precisely track the propagation, they still suffer difficulties in automatically analyzing the unexpected program behaviors caused by overflow.  Otherwise, they would be powerful enough to generate exploits for bugs, which is a problem far from being solved~\cite{avgerinos2014automatic}.

Figure~\ref{fig:bufferoverflow} demonstrates a buffer overflow example.  The program returns a \texttt{BOMB\_ENDING} if the value of \texttt{flag} equals one, which is unlikely because the value is zero and should remain unchanged without explicit modification.  However, the program has a buffer overflow bug.  It has a buffer (\texttt{buf}) of eight bytes and employs no boundary check when copying symbolic values to the buffer with \texttt{strcpy}.  We can change the value of \texttt{flag} to one leveraging the bug, \textit{e.g.,} when \texttt{symvar} is ``ANYSTRIN\textbackslash x01\textbackslash x00\textbackslash x00\textbackslash x00''.

\subsubsection{Parallel Executions}
Classic symbolic execution is effective for sequential programs.  We can draw an explicit CFG for sequential programs and let a symbolic execution engine traverse the CFG.  However, if a program processes symbolic variables in parallel, classic symbolic execution techniques would suffer problems.  Parallel programs can be undecidable because the execution order of parallel codes does not only depend on the program but may also depend on the execution context.  A parallel program may exhibit different behaviors even with the same test case.  This poses a problem for symbolic execution to generate test cases for triggering corresponding control flows.  If a symbolic execution tool directly ignores the parallel syntax or addresses the syntax improperly, errors would happen during $S_{inst}$ and $S_{sem}$.  

Figure~\ref{fig:paraprog} demonstrates an example with parallel codes.  The symbolic variable \texttt{i} is processed by another two additional threads in parallel, and the result is assigned to \texttt{j}.  Then the value of \texttt{j} determines the whether the program should return a \texttt{BOMB\_ENDING}.

To handle parallel codes, a symbolic execution tool has to interpret the semantics and track parallel executions, \textit{e.g.,} by introducing extra symbolic variables~\cite{farzan2013con2colic}.  However, such an approach may not be scalable because the possibility of parallel execution can be a large number.  In practice, there are several heuristic approaches to improve the efficiency.  For example, we may restrict the exploration time of concurrent regions with a threshold~\cite{farzan2013con2colic}; we may conduct symbolic execution with arbitrary contexts and convert multi-thread programs into equivalent sequential ones~\cite{bergan2014symbolic}; or we can prune unimportant paths leveraging some program codes, such as assertion~\cite{guo2015assertion}.

\subsubsection{Symbolic Memories}
Symbolic memory is a situation whereas symbolic variables serve as the offsets or pointers to retrieve values from the memory, such as array indexes.  To handle symbolic memories, a symbolic execution engine should take advantage of the memory layout for analysis.  For example, we can convert an array selection operation to a \texttt{switch/case} clause in which the number of possible cases equals the length of the array.  However, the number of possible combinations would grow exponentially when there are several such operations along a control flow.  In practice, a symbolic execution tool may directly employ the feature of array operations implemented by some constraint solvers, such as STP~\cite{ganesh2007decision} and Z3~\cite{de2008z3}.  It may also analyze the alignment of some pointers in advance, such as CUTE~\cite{sen2005cute}.  However, the power of pointer analysis is limited because the problem can be NP-hard or even undecidable for static analysis~\cite{landi1991pointer}.  If a symbolic execution tool cannot model symbolic memories properly, errors would occur during $S_{inst}$ and $S_{sem}$.

Figure~\ref{fig:symarray} demonstrates a sample of symbolic memories.  In this example, the symbolic variable \texttt{i} serves as an offset to retrieve an element from the array.  The retrieved element then determines whether the program returns a \texttt{BOMB\_ENDING}. 

\subsubsection{Contextual Symbolic Values}
The challenge is similar to symbolic memories but more complicated.  Other than retrieving values from the memory like symbolic memories, symbolic values can also serve as the parameters to retrieve values from the environment, such as loading the contents of a file pointed by symbolic values.  By default, this contextual information is unavailable to the program or process, and the analysis is complicated.  Moreover, since the contextual information can be changed any time without informing the program, the problem is undecidable.  A symbolic tool that does not support such operations would cause errors during $S_{inst}$ and $S_{sem}$.

Figure~\ref{fig:symval} is an example of contextual symbolic values.  If \texttt{symvar} points to an existed file on the local disk, the program would return a \texttt{BOMB\_ENDING}.

\subsubsection{Symbolic Jumps}
In general, symbolic execution only extracts constraint models when encountering conditional jumps, such as \texttt{var<0} in source codes, or \texttt{jle 0x400fda} in assembly codes.  However, we may also employ unconditional jumps to achieve the same effects as conditional jumps.  The idea is to jump to an address controlled by symbolic values.  If a symbolic execution engine is not tailored to handle the unconditional jumps, it would fail to extract corresponding constraint models and miss some available control flows.  Therefore, the challenge attacks the constraint modeling stage $S_{model}$.

Figure~\ref{fig:symjmp} is an example of symbolic jumps.  The program contains an array of function pointers, and each function returns an integer value.  The symbolic variable serves as an offset to determine which function should be called during execution.  If \texttt{f5()} is called, the program would return a \texttt{BOMB\_ENDING}.

\subsubsection{Floating-point Numbers}
A floating-point number ($f \in \mathbb{F}$) approximates a real number ($r \in \mathbb{R}$) with a fixed number of digits in the form of $f=sign*base^{exp}$.  For example, the 32-bit float type compliant to IEEE-754 has 1-bit for $sign$, 23-bit for $base$, and 8-bit for $exp$.  The representation is essential for computers, as the memory spaces are limited in comparison with the infinity of $\mathbb{R}$.  As a tradeoff, floating-point numbers only have limited precision, which makes some unsatisfiable constraints over $\mathbb{R}$ to be satisfied over $\mathbb{F}$ with a rounding mode.  In order to support reasoning over $\mathbb{F}$, a symbolic execution tool should consider such approximations when extracting and solving constraint models.  However, recent studies (\textit{e.g.,}~\cite{solovyev2015rigorous,quan2016hotspot,liew2017floating,liew2018symbolic}) show that there is still no silver bullet for the problem.  Floating-point numbers remains a challenge for symbolic execution tools, and the challenge attacks $S_{model}$.  

Figure~\ref{fig:floatpoint} demonstrates an example with floating-point operations.  Because we cannot represent $0.1$ with float type precisely, the first predicate \texttt{a != 1} is always true.  If the second condition \texttt{a == b} can be satisfied, the program would return a \texttt{BOMB\_ENDING}.  Therefore, one test case to returning a \texttt{BOMB\_ENDING} is \texttt{symvar} equals `7'.  

\subsubsection{Arithmetic Overflows}
Arithmetic overflow happens when the result of an arithmetic operation is outside the range of an integer type.  For example, the range of a 64-bit signed integer is $[-2^{64}, 2^{64}-1]$.  In this case, a constraint model (\textit{e.g.,} the result of a positive integer plus another positive integer is negative) may have no solutions over $\mathbb{R}$; but it can have solutions when we consider arithmetic overflow.  Handling such arithmetic overflow issues is not as difficult as previous challenges.  However, some preliminary symbolic execution tools may fail to consider these cases and suffer errors when extracting and solving the constraint models.

Figure~\ref{fig:intoverflow} shows a sample with an arithmetic overflow problem.  To meet the first condition \texttt{254748364 * i < 0}, \texttt{i} should be a negative value.  However, the second condition requires \texttt{i} to be a positive value.  Therefore, it has no solutions in the domain of real numbers.  But the conditions can be satisfied when \texttt{254748364 * i} exceeds the max value that the integer type can represent.

\subsection{Path-explosion Challenges}
Now we discuss three path-explosion challenges existed in small-size programs.

\subsubsection{External Function Calls}
Shared libraries, such as \texttt{libc} and \texttt{libm} (\textit{i.e.,} a math library), provide some basic function implementations to facilitate software development.  An efficient way to employ the functions is via dynamic linkage, which does not pack the function body to the program but only links with the functions dynamically when execution.  Therefore, such external functions do not enlarge the size of a program, but they can enlarge the code complexity in nature.

When an external function call is related to the propagation of symbolic values, the control flows within the function body should be analyzed by default.  There are two situations.  A simple situation is that the external function does not affect the program behaviors after executing it, such as simply printing symbolic values with \texttt{printf}.  In this case, we may ignore the path alternatives within the function.  However, if the function execution affects the follow-up program behaviors, we should not ignore them.  Otherwise, the symbolic execution would be based on a wrong assumption that the new test case generated for an alternative path can always trigger the same control flow within the external function.  If a small program contains several such function calls, the complexity of external functions may cause path explosion issues.  In practice, there are different strategies that symbolic execution tools may adopt with a trade-off between consistency and efficiency~\cite{chipounov2011s2e}.

Figure~\ref{fig:extfun} demonstrates a sample with an external function call.  It computes the sine of a symbolic value via an external function call (\textit{i.e.,} \texttt{sin}), and the result is used to determine whether the program should return a \texttt{BOMB\_ENDING}.

\subsubsection{Loops}
Loop statements, such as \texttt{for} and \texttt{while}, are widely employed in real-world programs.  Even a very small program with loops can include many or even an infinite number of paths.  By default, a symbolic execution tool should explore all available paths of a program, which can beyond the capability of the tool if there are too many paths.  In practice, a symbolic execution tool may employ a search strategy which favorites the unexplored branches on a program CFG~\cite{burnim2008heuristics,avgerinos2014enhancing}, or introduces new symbolic variables as the counters for each loop~\cite{saxena2009loop}.  Because loop can incur numerous paths, we can hardly have a perfect solution for this problem.

Figure~\ref{fig:loop} shows a sample with a loop.  The loop function is implemented with the Collaz conjecture~\cite{lagarias1985}.  No matter what is the initial value of \texttt{i}, the loop will terminate with \texttt{j} equals 1.

\subsubsection{Crypto Functions}
Crypto functions generally involve some computationally complex problems to ensure security.  For a hash function, the complexity guarantees that adversaries cannot efficiently compute the plaintext of a hash value.  For a symmetric encryption function, it promises that one cannot efficiently compute the key when given several pairs of plaintext and ciphertext.  Therefore, such programs should also be resistant to symbolic execution attacks.  From a program analysis view, the number of possible control paths for the crypto functions can be substantial.  For example, the body of the SHA1 algorithm~\cite{eastlake2001us} is a loop that iterates 80 rounds, and each round contains several bit-level operations.  

Figure~\ref{fig:cryptofun} demonstrates a code snippet which employs a SHA1 function~\cite{eastlake2001us}.  If the hash result of the symbolic value is equivalent to a predefined value, the program would return a \texttt{BOMB\_ENDING}.  However, it is difficult since SHA1 cannot be reversely calculated.

In general, symbolic execution tools cannot handle such crypto programs.  Malware may employ the technique to deter symbolic execution-based program analysis~\cite{sharif2008impeding}.  When analyzing programs with crypto functions, a common way is to avoid exploring the function internals (\textit{e.g.,}\cite{wang2010taintscope,corin2011efficient}).  For example, TaintScope~\cite{wang2010taintscope} first discriminates the symbolic variables corresponding to crypto functions from other variables, and then it employs a fuzzy-based approach to search solutions for such symbolic variables rather than solving the problem via symbolic reasoning.

So far, we have discussed 12 different challenges in total.  Note that we do not intend to propose a complete list of challenges for symbolic execution.  Instead, we collect all the challenging issues that have been mentioned in the literature and systematically analyze them.  This analysis is essential for us to design the the dataset of logic bombs in Section~\ref{sec:dataset}.

\section{Benchmark Methodology} \label{sec:method}
In this section, we introduce our methodology and a framework to benchmark the capability of real-world symbolic execution tools.

\subsection{Objective and Challenges}
Before describing our approach, we first discuss our design goal and the challenges to overcome.  

This work aims to design an approach that can benchmark the capabilities of symbolic execution tools.  Our purpose is critical and valid in several aspects.  As we have discussed, some challenging issues are only engineering issues, such as arithmetic overflows.  With enough engineering effort, a symbolic execution tool should be able to handle these issues.  On the other hand, some challenges are hard from a theoretical view, such as loops.  But some heuristic approaches can tackle certain easy cases.  Symbolic execution tools may adopt different heuristics and demonstrate different capabilities in handling them.  Therefore, it is worth benchmarking their performances in handling particular challenging issues.  Developers generally do not provide much information about the limitations of their tools to users.

A useful benchmark approach should be accurate and efficient.  However, it is challenging to benchmark symbolic execution tools accurately and efficiently.  Firstly, a real program contains many instructions or lines of codes.  When a symbolic execution failure happens, locating the root cause requires much domain knowledge and effort.  Since errors may propagate, it is challenging to conjecture whether a symbolic execution tool fails in handling a particular issue.  Secondly, the symbolic execution itself is inefficient.  Benchmarking a symbolic execution tool generally implies performing several designated symbolic execution tasks, which would be time-consuming.  Note that existing symbolic execution papers (\textit{e.g.,}~\cite{cadar2008klee,kuznetsov2012efficient,shoshitaishvili2016state,hasabnis2016extracting}) generally evaluate the performance of their tools by conducting symbolic execution experiments with real programs.  The process usually takes several hours or even days.  They demonstrate the effectiveness of their work using the achieved code coverage and number of bugs detected, and they do not analyze the root causes of uncovered codes.

\subsection{Approach based on Logic Bombs}
To tackle the challenges of benchmark concerning accuracy and efficiency, we propose an approach based on logic bombs.  Below, we discuss our detailed design.

\subsubsection{Evaluation with Logic Bombs}

\begin{algorithm} [t!b]
\caption{Method to design evaluation samples.}
\label{alg:method}
\small
\SetKwProg{Fn}{Function}{}{}
\tcp{Create a function with a symbolic variable}
\Fn{LogicBomb($symvar$)}{
    \tcp{$symvar2$ is a value computed from a challenging problem related to $symvar$}
    $symvar2 \gets$ Challenge($symvar$)\;
    \tcp{If $symvar2$ satisfies a condition}
    \If{Condition($symvar2$)}{
        \tcp{Trigger the bomb}
        Bomb()\;
    }
}
\end{algorithm}

A logic bomb is a code snippet that can only be executed when certain conditions are met.  To evaluate whether a symbolic execution tool can handle a challenge, we can design a logic bomb guarded by a particular issue with the challenge.  Then we can perform symbolic execution on the program which embeds the logic bomb.  If a symbolic execution tool can generate a test case that can trigger the logic bomb, it indicates the tool can handle the challenging issue, or \textit{vice versa}.

Algorithm~\ref{alg:method} demonstrates a general framework to design such logic bombs.  It includes four steps: the first step is to create a function with a parameter $symvar$ as the symbolic variable; the second step is to design a challenging problem related to the symbolic variable and save the result to another variable $symvar2$; the third step is to design a condition related to the new variable $symvar2$; the final step is to design a bomb (\textit{e.g.,} return a specific value) which indicates the condition has been satisfied.  Note that because the value of $symvar2$ is propagated from $symvar$, $symvar2$ is also a symbolic variable and should be considered in the symbolic analysis process.

The magic of the logic bomb idea enables us to make the evaluation much precise and efficient.  We can create several such small programs; each contains only a challenging issue and a logic bomb that tells the evaluation result.  Because the object programs for symbolic execution are small, we can easily avoid unexpected issues that may also cause failures via a careful design.  Also because the programs are small, performing symbolic execution on them generally requires a short time.  For the programs that unavoidably incur path explosion issues, we can restrict the symbolic execution time either by controlling the problem complexity or by employing a timeout setting.

\subsubsection{Logic Bomb Dataset} \label{sec:dataset}

\begin{figure*}[t]
\centering
\includegraphics[width=0.96\textwidth]{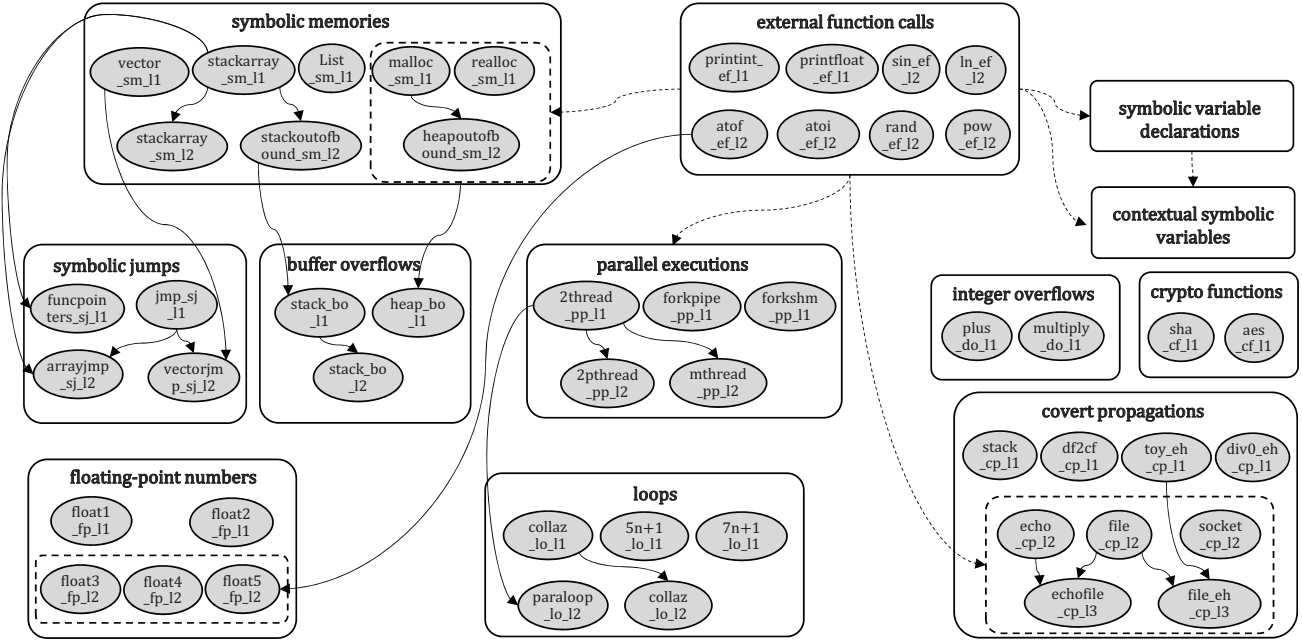}
\caption{The challenge propagation relationship among our dataset of logic bombs.  A solid line means a logic bomb contains a similar problem defined in another logic bomb; a dashed line means a challenge may affect other logic bombs.}
\label{fig:dataset}
\end{figure*}

Following Algorithm~\ref{alg:method}, we have designed a dataset of logic bombs to evaluate the capability of symbolic execution tools.  We have already shown several samples in Figure~\ref{fig:examples}.  Our full dataset is available on GitHub\footnote{https://github.com/hxuhack/logic\_bombs}.  The dataset contains over 60 logic bombs for 64-bit Linux platform, which covers all the challenges discussed in Section~\ref{sec:challenges}.  For each challenge, we implemented several logic bombs.  Either each bomb involves a unique challenging issue (\textit{e.g.,} covert propagation via file write/read or via system calls), or introduces a problem with a different complexity setting (\textit{e.g.,} one-leveled arrays or two-leveled arrays).

When designing the logic bombs, we carefully avoid trivial test cases (\textit{e.g.,} \texttt{\textbackslash x00}) that can trigger the bombs.  Moreover, we try to employ straightforward implementations, and we hope to ensure that the results would not be affected by other unexpected failures.  For example, we avoid using \texttt{atoi} to convert \texttt{argv[1]} to integers because some tools cannot support \texttt{atoi}.  However, fully avoiding external function calls is impossible for some logic bombs.  For example, we should employ external function calls to create threads when designing parallel codes.  Surely that if a symbolic execution tool cannot handle external functions, the result might be affected.  To tackle the interference of challenges, we draw a challenge propagation chart among the logic bombs as shown in Figure~\ref{fig:dataset}.  There are two kinds of challenge propagation relationships: $should$ in solid lines, and $may$ in dashed lines.  A $should$ relationship means a logic bomb contains a similar challenging issue in another logic bomb; if a tool cannot solve the precedent logic bomb, it should not be able to solve the later one.  For example, the \texttt{stackarray\_sm\_l1} is precedent to \texttt{stackarray\_sm\_l2}.  A $may$ relationship means a challenge type may be a precedent to other logic bombs, but it is not the determinant one.  For example, a parallel program generally involves external function calls.  However, although a tool is unable to solve the external functions well, it might be able to solve some logic bombs with parallel issues as sequential programs.

\subsection{Automated Benchmark Framework}\label{sec:framework}
\begin{figure}[t!b]
\centering
\includegraphics[width=0.46\textwidth]{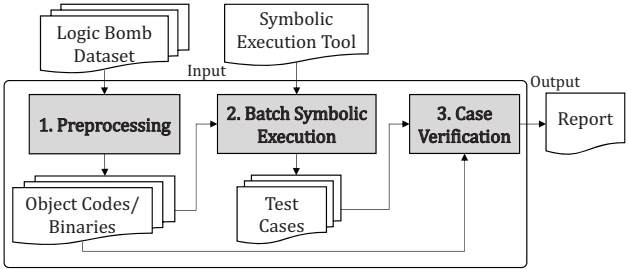}
\caption{A framework to benchmark symbolic execution tools.}
\label{fig:framework}
\end{figure}

Base on the evaluation idea with logic bombs, we design a benchmark framework as shown in Figure~\ref{fig:framework}.  The framework inputs a dataset of carefully designed logic bombs and outputs the benchmark result for a particular symbolic execution tool.  There are three critical steps in the framework: dataset preprocessing, batch symbolic execution, and case verification.

In the preprocessing step, we parse the logic bombs and compile them into object codes or binaries such that a target symbolic execution tool can process them.  The parsing process pads each code snippet of a logic bomb with a main function and makes it a self-contained program.  By default, we employ \texttt{argv[1]} as the symbolic variables.  If a target symbolic execution tool requires adding extra instructions to launch tasks, the parser should add such required instructions automatically.  For example, we can add symbolic variable declaration codes when benchmarking KLEE.  The compilation process compiles the processed source codes into binaries or other formats that a target symbolic execution tool supports.  Symbolic execution is generally performed based on intermediate codes.  When benchmarking source-code-based symbolic execution tools such as KLEE, we have to compile the source codes into the supported intermediate codes.  When benchmarking binary-code-based symbolic execution tools, we can directly compile them into binaries, and the tool will lift binary codes into intermediate codes automatically.

In the second step, we direct a symbolic execution tool to analyze the compiled logic bombs in a batch mode.  This step outputs a set of test cases for each program.  Some dynamic symbolic execution tools (\textit{e.g.,} Triton) can directly tell which test case can trigger a logic bomb during runtime.  However, other static symbolic execution tools may only output test cases by default, and we need to replay the generated test cases to examine the results further.  Besides, some tools may falsely report that a test case can trigger the logic bomb.  Therefore, we need a third step to verify the test cases. 

In the third step, we replay the test cases with the corresponding programs of logic bombs.  If a logic bomb can be triggered, it indicates that the challenging case is solved by the tool.  Finally, we can generate a benchmark report based on the case verification results. 

\section{Experimental Study} \label{sec:experiment}
In this section, we conduct an experimental study to demonstrate the effectiveness of our benchmark approach.  Below, we discuss the experimental setting and results.

\subsection{Experimental Setting}
We choose three popular symbolic execution tools for benchmark: KLEE~\cite{cadar2008klee}, Angr~\cite{shoshitaishvili2016state}, and Triton~\cite{saudel2015triton}.  Because our dataset of logic bombs are written in C/C++, we only choose symbolic execution tools for C/C++ programs or binaries.  The three tools are all released as open source and have a high community impact.  Moreover, they adopt different implementation techniques for symbolic execution.  By supporting variant tools, we show that our approach is compatible with different symbolic execution implementations.

KLEE\cite{cadar2008klee} is a static symbolic execution tool implemented based on LLVM~\cite{lattner2004llvm}.  It requires program source codes to perform symbolic execution.  By default, our benchmark script employs a \texttt{klee\_make\_symbolic} function to declare the symbolic variables of logic bombs in the source-code level.  Then, it compiles the source codes into intermediate codes for symbolic execution.  The symbolic execution process outputs a set of test cases, and our script finally examines the test cases by replaying them with the binaries.  The whole process is automated with our benchmark script.  The version of KLEE we benchmark is 1.3.0.  Note that because our experiment does not intend to find the best tool for particular challenges, so we do not consider the patches provided by other parties before they are merged into the master branch.

Triton~\cite{saudel2015triton} is a dynamic symbolic execution tool based on binaries.  It automatically accepts symbolic variables from the standard input.  During symbolic execution, Triton firstly runs the programs with concrete values and leverages Intel PinTool~\cite{luk2005pin} to trace related instructions, then it lifts the traced instructions into the SSA (single static assignment) form and performs symbolic analysis.  If there are alternative paths found in the trace, Triton generates new test cases via symbolic reasoning and employs them as the concrete values in the following rounds of concrete execution.  This symbolic execution process goes on until no alternative path can be found.  The version of Triton we adopted is which released on Jul 6, 2017, on GitHub.

Angr~\cite{shoshitaishvili2016state} is also a tool for binaries but employs different implementations.  Before performing any symbolic analysis, Angr firstly lifts the binary program into VEX IR~\cite{nethercote2004dynamic}.  Then it employs a symbolic analysis engine (\textit{i.e.,} SimuVEX) to analyze the program based on the IR.  Angr does not provide ready-to-use symbolic execution script for users but only some APIs.  Therefore, we have to implement our own symbolic execution script for Angr.  Our script collects all the paths to the CFG leaf nodes and then solves the corresponding path constraints.  Angr provides all the critical features via APIs, and we only assemble them.  Finally, we check whether the generated test cases can trigger the logic bombs.  In our experiment, we employ Angr with version 7.7.9.21.

Note that our benchmark scripts for these tools all follow the framework proposed in Figure~\ref{fig:framework}.  During the experiment, we employ our logic bomb dataset for evaluation.  A tool can pass a test only if the generated solution can correctly trigger a logic bomb.  We finally report which logic bombs can be triggered by the tools.

We conduct our experiments on an Ubuntu 14.04 X86\_64 system with Intel i5 CPU and 8G RAM.  Because some symbolic execution tasks may take very long time, our tool allows users to configure a timeout threshold which ensures benchmark efficiency.  However, the timeout mechanism may incur some false results if it is too short.  To mitigate the side effects, we adopt two timeout settings (60 seconds and 300 seconds) for each tool.  In this way, we can observe the influence of the timeout settings and decide whether we should conduct more experiments with an increased timeout value.  
 
\subsection{Benchmark Results}

\begin{table*}[h]
\centering
\caption{Experimental results on benchmarking three symbolic execution tools (KLEE, Triton, and Angr) in handling our logic bombs.  Pass means the tool has successfully triggered the bomb; fail means the tool cannot find test cases to trigger the bomb; timeout means the tool cannot find test cases to trigger the bomb within a given period of time.  For each tool, we adopt two timeout settings: 60 seconds and 300 seconds.}
\label{tab:result}
\newcommand{\tabincell}[2]{\begin{tabular}{@{}#1@{}}#2\end{tabular}}
\renewcommand{\multirowsetup}{\centering} 
\tabcolsep=0.2cm
\begin{tabular}{|c|c|c|c|c|c|c|c|}
\hline
\multirow{2}{*}{\textbf{Challenge}} & \multirow{2}{*}{\textbf{\tabincell{c}{Case ID}}} & \multicolumn{2}{|c|}{\textbf{KLEE}} & \multicolumn{2}{|c|}{\textbf{Triton}} & \multicolumn{2}{|c|}{\textbf{Angr}} \\
\cline{3-8}
 & & t = 60s & t = 300s & t = 60s & t = 300s & t = 60s & t = 300s  \\
\hline
\multirow{9}{*}{\tabincell{c}{Covert Propagations}} & df2cf\_cp & pass & pass & fail & fail & pass & pass \\
\cline{2-8}
& echo\_cp & fail & fail & timeout & timeout & timeout & timeout \\
\cline{2-8}
& echofile\_cp & fail & fail & fail & fail & timeout & timeout \\
\cline{2-8}
& file\_cp & fail & fail & timeout & timeout & fail & fail \\
\cline{2-8}
& socket\_cp & fail & fail & fail & fail & fail & fail \\
\cline{2-8}
& stack\_cp & fail & fail & pass & pass & pass & pass \\
\cline{2-8}
& file\_eh\_cp & fail & fail & fail & fail & timeout & pass \\
\cline{2-8}
& div0\_eh\_cp & fail & fail & fail & fail & timeout & pass \\
\cline{2-8}
& file\_eh\_cp & fail & fail & fail & fail & timeout & fail \\
\hline
\multirow{3}{*}{\tabincell{c}{Buffer Overflows}}  & stack\_bo\_l1 & fail & fail & fail & fail & pass & pass \\
\cline{2-8}
 & heap\_bo\_l1 & fail & fail & fail & fail & fail & fail \\
 \cline{2-8}
 & stack\_bo\_l2 & fail & fail & fail & fail & fail & fail \\
\hline
\multirow{9}{*}{\tabincell{c}{Symbolic Memories}} & malloc\_sm\_l1 & pass & pass & timeout & fail & pass & pass \\
 \cline{2-8}
 & realloc\_sm\_l1 & pass & pass & fail & fail & pass & pass \\
  \cline{2-8}
 & stackarray\_sm\_l1 & pass & pass & fail & fail & pass & pass \\
  \cline{2-8}
 & list\_sm\_l1 & fail & fail & fail & fail & timeout & pass \\
  \cline{2-8}
 & vector\_sm\_l1 & fail & fail & fail & fail & timeout & pass \\
  \cline{2-8}
 & stackarray\_sm\_l2 & pass & pass & fail & fail & fail & fail \\
  \cline{2-8}
 & stackoutofbound\_sm\_l2 & pass & pass & fail & fail & pass & pass \\
  \cline{2-8}
 & heapoutofbound\_sm\_l2 & fail & fail & timeout & fail & pass & pass \\
\hline
\multirow{3}{*}{\tabincell{c}{Symbolic Jumps}} & funcpointer\_sj\_l1 & pass & pass & fail & fail & fail & fail \\
  \cline{2-8}
& jmp\_sj\_l1 & fail & fail & fail & fail & pass & pass \\
  \cline{2-8}
& arrayjmp\_sj\_l2 & fail & fail & fail & fail & fail & fail \\
  \cline{2-8}
& vectorjmp\_sj\_l2 & fail & fail & fail & fail & timeout & pass \\
\hline
\multirow{5}{*}{\tabincell{c}{Floating-point Numbers}} & float1\_fp\_l1 & fail & fail & fail & fail & pass & pass \\
  \cline{2-8}
 & float2\_fp\_l1 & fail & fail & fail & fail & pass & pass \\
 \cline{2-8}
 & float3\_fp\_l2 & fail & fail & fail & fail & timeout & timeout \\
 \cline{2-8}
 & float4\_fp\_l2 & fail & fail & fail & fail & timeout & timeout \\
 \cline{2-8}
 & float5\_fp\_l2 & fail & fail & fail & fail & timeout & timeout \\
\hline
\multirow{2}{*}{\tabincell{c}{Arithmetic Overflows}} & plus\_do & pass & pass & pass & pass & pass & pass \\
 \cline{2-8}
& multiply\_do & pass & pass & fail & fail & pass & pass \\
\hline
\multirow{8}{*}{\tabincell{c}{External Function Calls}} & printint\_ef\_l1 & fail & fail & pass & pass & pass & pass \\
 \cline{2-8}
& printfloat\_ef\_l1 & fail & fail & fail & fail & fail & fail \\
 \cline{2-8}
& atoi\_ef\_l2 & fail & fail & fail & fail & pass & pass \\
 \cline{2-8}
& atof\_ef\_l2 & fail & fail & fail & fail & timeout & timeout \\
 \cline{2-8}
& ln\_ef\_l2 &fail & fail & fail & fail & timeout & fail \\
 \cline{2-8}
& pow\_ef\_l2  &fail & fail & fail & fail & pass & pass \\
 \cline{2-8}
& rand\_ef\_l2  &fail & fail & timeout & timeout & fail & fail \\
 \cline{2-8}
& sin\_ef\_l2  &fail & fail & fail & fail & timeout & timeout \\
\hline
\tabincell{c}{Symbolic Variable Declarations} & \multicolumn{7}{|c|}{7 cases, all fail} \\
\hline
\tabincell{c}{Contextual Symbolic Values} & \multicolumn{7}{|c|}{4 cases, all fail}\\
\hline
\tabincell{c}{Parallel Executions} & \multicolumn{7}{|c|}{5 cases, all fail}\\
\hline
Loops & \multicolumn{7}{|c|}{5 cases, all fail}\\
\hline
Crypto Functions & \multicolumn{7}{|c|}{2 cases, all fail}\\
\hline
\textbf{pass \#} & 62 cases & 9 & 9 & 3 & 3 & 16 & 21 \\
\hline
\end{tabular}
\end{table*}

\subsubsection{Result Overview}
Table~\ref{tab:result} demonstrates our experimental results.  We label the results with three options: pass, fail, and timeout.  While `pass' and `fail' imply the symbolic execution has finished, `timeout' implies our benchmark script has terminated the symbolic execution process when a timeout threshold is triggered.  

From the results, we can observe that Angr has achieved the best performance with 21 cases solved when the timeout is 300 seconds.  Comparatively, it only solved 16 cases when the timeout is 60 seconds.  KLEE solved nine cases and the result remains the same with different timeout settings.  Triton performs much worse with three cases solved.  To further verify the correctness of our benchmark results, we compare our experimental results with the previously declared challenge propagation relationships in Figure~\ref{fig:dataset}.  We find the results are all consistent.  It justifies that our dataset can distinguish the capability of different symbolic-execution tools accurately and effectively.

The efficiency of our benchmark approach largely depends on the timeout setting.  Note that Table~\ref{tab:result} includes some timeout results, and they account for most of our experimental time.  Although we try to keep each logic bomb as succinct as possible, our dataset still contains some complex problems or path explosion issues unavoidable.  When the timeout value is 60 seconds, our benchmark process for each tool takes only dozens of minutes.  When extending the timeout value to 300 seconds, the benchmark takes a bit longer time.  However, the benefit is not very obvious, and only Angr can solve 5 more cases.  Can the result get further improved with more time?  We have tried another group of experiments with 1800 seconds timeout and the results remain unchanged.  Therefore, 300 seconds should be a marginal timeout setting for our benchmark experiment.  Considering that symbolic execution is computationally expensive, which may take several hours or even several days to test a program, our benchmark process is very efficient.  We may further improve the efficiency by employing a parallel mode, such as assigning each process several logic bombs.

\subsubsection{Case Study} \label{sec:case_study}
Now we discuss the detailed benchmark results for each challenge.  Firstly, there are several challenges that none of the tools can trigger even one logic bomb, including symbolic variable declarations, parallel executions, contextual symbolic values, loops, and crypto functions.  For symbolic variable declaration challenge, because all the tools cannot recognize the expected symbolic variables automatically, they fail in modeling the conditions to trigger the logic bombs.  The challenges of contextual symbolic values and crypto functions involve tough problems, and it can be expected that all the tools fail in handling them.  However, it is a bit surprising that none of the tools can handle parallel executions and loops.

\begin{figure*}[t!b]
	\hspace{1cm}
    \subfigure[Source codes.]{
    \label{fig:div0}
    \includegraphics[width=0.3\textwidth]{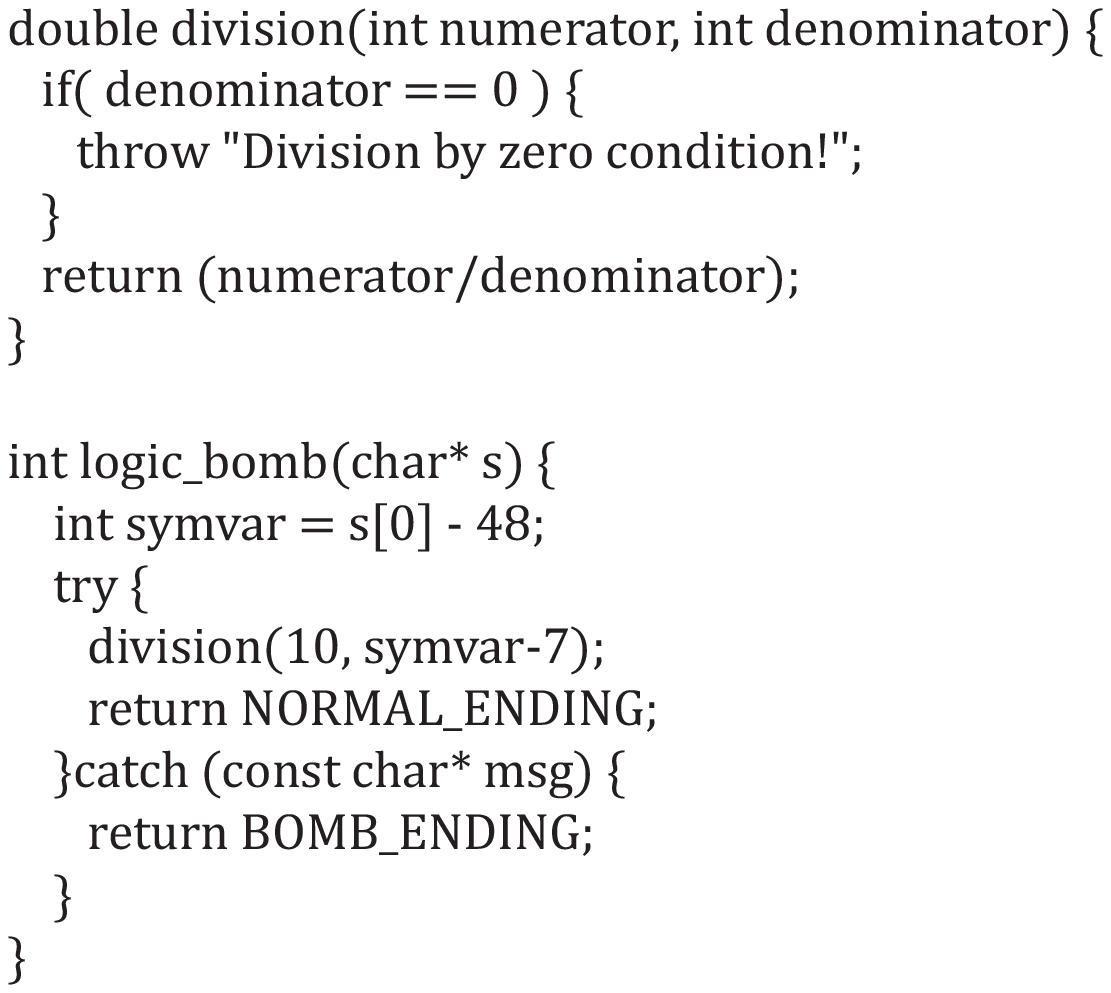}}
    \hspace{1cm}
    \subfigure[Assembly codes.]{
    \label{fig:div0_as}
    \includegraphics[width=0.45\textwidth]{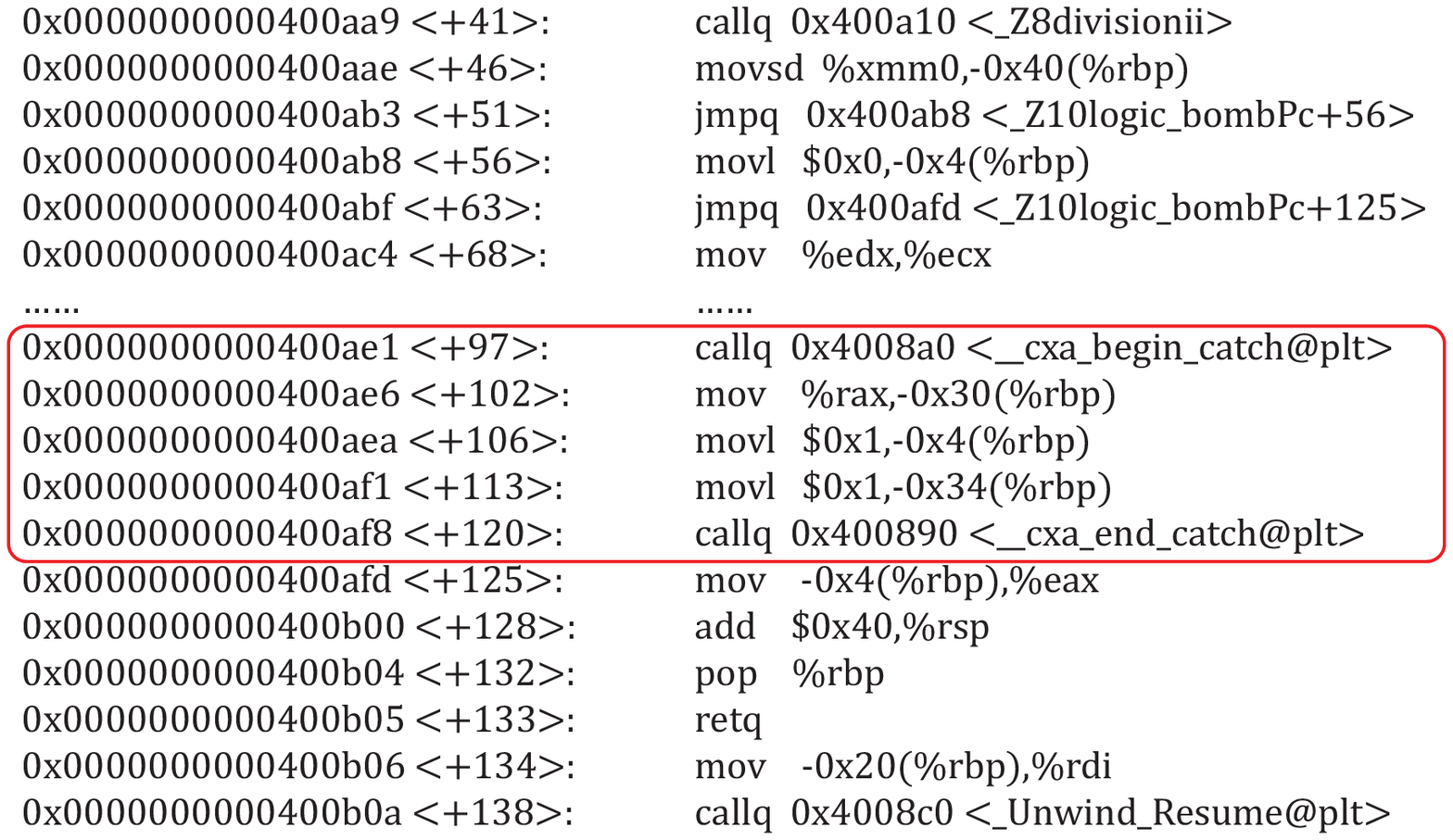}}
    \caption{An exemplary program that raises an exception when divided by zero.  The assembly codes demonstrates how the \texttt{try/catch} mechanism works in low level.}
    \label{fig:exception}
\end{figure*}

\begin{figure*}[th!b]
    \begin{minipage}[c][11cm][t]{.5\textwidth}
    \subfigure[Source codes.]{
    \label{fig:l2array}
    \includegraphics[width=0.8\textwidth]{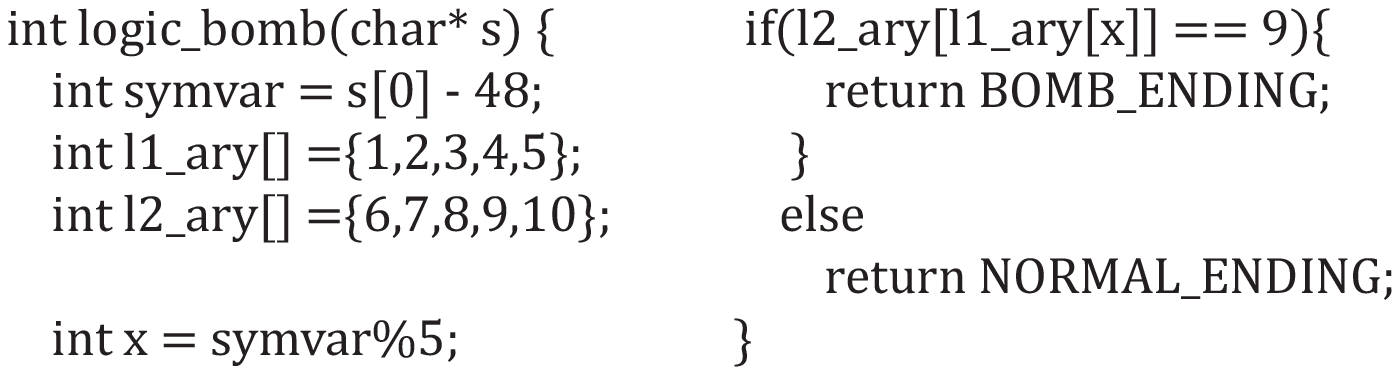}}\\
    \\
    \\
    \subfigure[Memory layout after array initialization.]{
    \label{fig:l2array_mem}
    \includegraphics[width=0.8\textwidth]{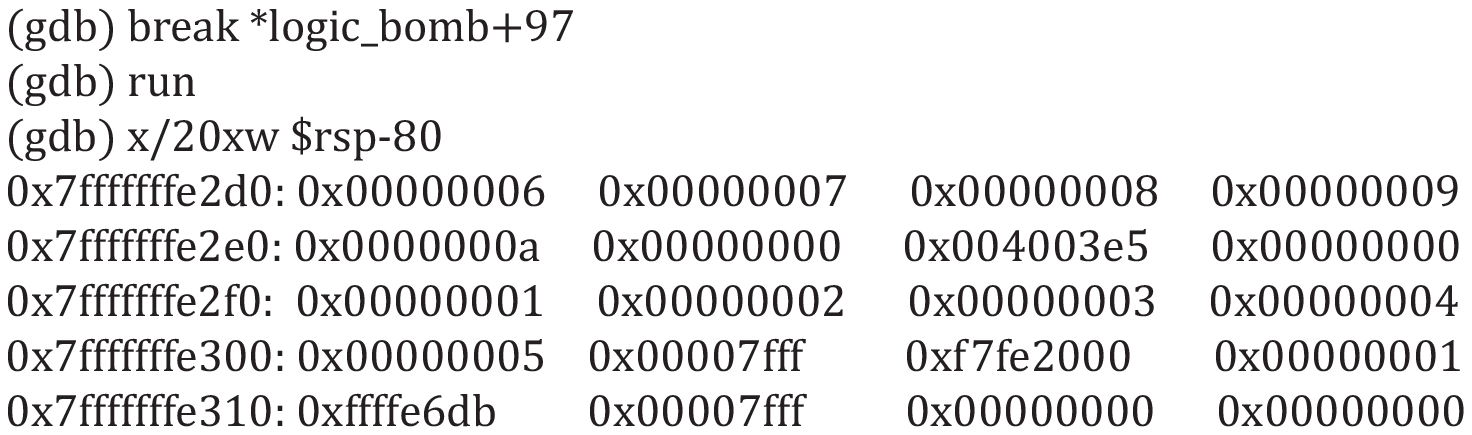}}
    \end{minipage}
    \begin{minipage}[c][11cm][t]{.5\textwidth}
    \centering
    \subfigure[Assembly codes.]{
    \label{fig:l2array_as}
    \includegraphics[width=0.8\textwidth]{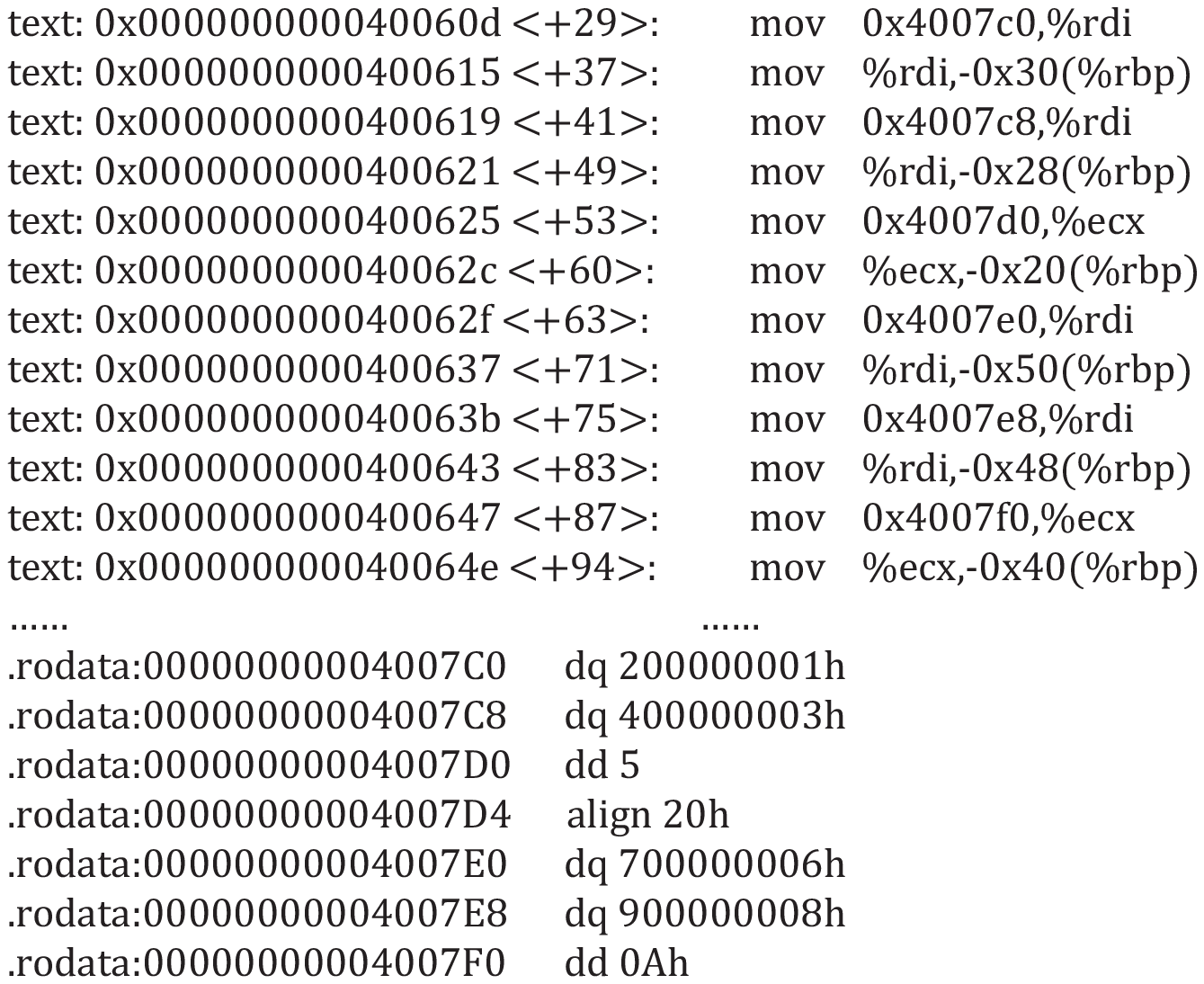}}
    \end{minipage}
    \vspace{-5cm}
    \caption{An exemplary program that demonstrates how the stack works with arrays.  There is no information about the size of each array left in assembly codes.}
\end{figure*}

\textit{Covert Propagations}:  Angr have passed four test cases, \texttt{df2cf\_cp}, \texttt{stack\_cp}, and two exception handling cases.  \texttt{df2cf\_cp} propagates the symbolic values indirectly by substituting a data assignment operation with equivalent control-flow operations.  KLEE also solved the case, but Triton failed.  \texttt{stack\_cp} propagates symbolic values via direct assembly instructions \texttt{push} and \texttt{pop}.  Only KLEE failed the test because it is a source-based analysis tool which does not support assembly codes.  Besides, Angr has also passed two test cases that propagate symbolic values via the C++ exception handling mechanism.  Not only this exception handling mechanism can be covert for some source-code-based analysis tools, such as KLEE, but also it can be covert for binary-code-based symbolic execution tools, such as Triton.  We further break down the details of an exception handling program in Figure~\ref{fig:exception}.  As shown in the box region of Figure~\ref{fig:div0_as}, the mechanism relies on two function calls, which might be the problem that fails Triton.  All the tools failed other covert propagation cases that propagate values via file read/write, echo, socket, \textit{etc}.

\textit{Buffer Overflows}:  Only Angr has solved one easy buffer overflow problem \texttt{stack\_bo\_l1}.  The case has a simple stack overflow issue.  Its solution requires modifying the value of the stack that might be illegal.  However, Angr cannot solve the heap overflow issue \texttt{heap\_bo\_l1}.  It also failed on another harder stack overflow issue \texttt{stack\_bo\_l2}, which requires composing sophisticated payload, such as employing return-oriented programming methods~\cite{roemer2012return}.  We are surprised that Triton failed all the tests because binary-code-based symbolic execution tools should be resilient to buffer overflows in nature.

\textit{Symbolic Memories}:  The results show that Triton does not support symbolic memory, but KLEE and Angr provide very good support.  Angr has solved seven cases out of eight.  It only failed in handling the case (Figure~\ref{fig:l2array}) with a two-leveled array \texttt{stackarray\_sm\_l2}.  Also, it implies that when there are multi-leveled pointers, Angr would fail.  Figure~\ref{fig:l2array_as} demonstrates the assembly codes that initialize the arrays, and Figure~\ref{fig:l2array_mem} demonstrates the stack layout after initialization.  We can observe that the information about array size or boundary does not exist in assembly codes.  This justifies why binary-code-based symbolic execution tools do not suffer problems when a challenge requires an out-of-boundary access, \textit{e.g.,} \texttt{stackoutofbound\_sm\_l2}.  In comparison, KLEE can solve the two-leveled array problem because it is based on STP~\cite{ganesh2007decision}, which is designed for solving such problems related to arrays.  However, KLEE does not support C++, so it failed the problems with vectors and lists.

\textit{Symbolic Jumps}:  Since symbolic jump demonstrates no explicit conditional branches in the CFG, it should be a hard problem for symbolic execution.  However, KLEE and Angr are not likely to be affected much by the trick.  KLEE has tackled the problem with an array of function pointers \texttt{funcpointer\_sj\_l1}.  It failed the other test cases because they employ an assembly instruction \texttt{jmp}, which KLEE does not support.  Angr successfully handled two cases with assembly \texttt{jmp}, but it failed \texttt{funcpointer\_sj\_l1}.

\textit{Floating-point Numbers}:  The results show KLEE and Triton do not support floating-point operations, and Angr can support some.  During our test, Triton directly reported that it cannot interpret such floating-point instructions.  Angr has solved two out of the five designated cases.  The two passed cases are easier ones, which only require integer values as the solution.  All the failed cases require decimal values as the solution, and they employ the \texttt{atof} function to convert \texttt{argv[1]} to decimals.  Since Angr has also failed the test in handling \texttt{atof} in \texttt{ atof\_ef\_l2}, the failures are likely to be caused by the \texttt{atof} function.

\textit{Arithmetic Overflows}:  Arithmetic overflow is not a very hard problem, and it only requires symbolic execution tools to handle such cases carefully.  In our test, KLEE and Angr have solved all the cases.  However, Triton failed in handling the integer overflow case in Figure~\ref{fig:intoverflow}.  The result shows there is still much room for Triton to improve for this problem.

\textit{External Function Calls}:  In this group of logic bombs, each case only contains one external function call.  However, the result is very disappointing.  Triton only passed a very simple case that print out (with \texttt{printf}) a symbolic value of integer type.  It does not even support printing out floating-point values.  Angr has solved the \texttt{printf} cases and two more complicated cases, \texttt{atoi\_ef\_l2} and \texttt{pow\_ef\_l2}.  It cannot support \texttt{atof\_ef\_l2} and other cases.  The results show that we should be cautious when designing logic bombs.  Even when involving straightforward external function calls, the results could be affected.

\section{Conclusion} \label{sec:conclusion}
This work proposes an approach to benchmark the capability symbolic execution tools in handling particular challenges.  To this end, we studied the taxonomy of challenges faced by symbolic execution tools, including nine symbolic-reasoning challenges and three path-explosion challenges.  Such a study is essential for us to design the benchmark dataset.  Then we proposed a promising benchmark approach based on logic bombs.  The idea is to design logic bombs that can only be triggered if a symbolic execution tool solves specific challenging issues.  By making the programs of logic bombs as small as possible, we can speed up the benchmark process; and by making them as straightforward as possible, we can avoid unexpected reasons that may affect the benchmark results.  In this way, our benchmark approach is both accurate and efficient.  Following the idea, we implemented a dataset of logic bombs and a prototype benchmark framework which automates the benchmark process.  Then, we conducted real-world experiments on three symbolic execution tools.  Experimental results show that the benchmark process for each tool generally takes dozens of minutes.  Angr achieved the best benchmark results with 21 cases solved, KLEE solved nine, and Triton only solved three.  These results justify the value of a third-party benchmark toolset for symbolic execution tools.  Finally, we released our dataset as open source on GitHub for public usage.  We hope it would serve as an essential tool for the community to benchmark symbolic execution tools and could facilitate the development of more comprehensive symbolic execution techniques. 

\ifCLASSOPTIONcompsoc
  \section*{Acknowledgments}
\else
  \section*{Acknowledgment}
\fi

This work was supported by the National Natural Science Foundation of China (Project
Nos. 61672164,61332010), and 2015 Microsoft Research Asia Collaborative Research Program (Project No. FY16-RES-THEME-005).

\ifCLASSOPTIONcaptionsoff
  \newpage
\fi

\bibliographystyle{IEEEtran}
\bibliography{tdsc}

\end{document}